\documentclass[nofootinbib,prl,reprint,superscriptaddress,preprintnumbers]{revtex4-2}

\makeatletter 
    
\renewcommand\onecolumngrid{
\do@columngrid{one}{\@ne}%
\def\set@footnotewidth{\onecolumngrid}
\def\footnoterule{\kern-6pt\hrule width 1.5in\kern6pt}%
}

\renewcommand\twocolumngrid{
        \def\footnoterule{
        \dimen@\skip\footins\divide\dimen@\thr@@
        \kern-\dimen@\hrule width.5in\kern\dimen@}
        \do@columngrid{mlt}{\tw@}
}%

\makeatother    

\usepackage[bottom]{footmisc}
\usepackage{amsmath,amssymb,marginnote}
\usepackage{multirow} 
\usepackage{booktabs} 
\usepackage{graphicx,subfigure}
\usepackage{dcolumn}
\usepackage{bm}
\usepackage[mathlines]{lineno}
\usepackage[usenames]{color}
\usepackage{xcolor}
\usepackage[colorlinks=true, allcolors=blue]{hyperref}
\usepackage{comment}

\usepackage{mathrsfs}

\begin{document}

\preprint{LA-UR-25-29100}

\title{Dynamical equilibria of fast neutrino flavor conversion}

\author{Jiabao Liu}
\affiliation{Department of Physics and Applied Physics, School of Advanced Science \& Engineering, Waseda University, Tokyo 169-8555, Japan}
\author{Lucas Johns}
\affiliation{Theoretical Division, Los Alamos National Laboratory, Los Alamos, NM 87545, USA}
\author{Hiroki Nagakura}
\affiliation{Division of Science, National Astronomical Observatory of Japan, 2-21-1 Osawa, Mitaka, Tokyo 181-8588, Japan}
\author{Masamichi Zaizen}
\affiliation{Department of Earth Science and Astronomy, The University of Tokyo, Tokyo 153-8902, Japan}
\author{Shoichi Yamada}
\affiliation{Department of Physics, School of Advanced Science \& Engineering, Waseda University, Tokyo 169-8555, Japan}
\affiliation{Research Institute for Science and Engineering, Waseda University, Tokyo 169-8555, Japan}

\begin{abstract}
Dense neutrino systems, which display collectivity mediated by the weak interaction, have deep parallels with mean-field kinetic systems governed by other fundamental forces. We identify analogues in fast flavor conversion (FFC) of some time-honored nonlinear phenomena in plasmas and self-gravitating systems. We focus in particular on nonlinear Landau damping and collisionless equilibria, which are likely important pieces of the unsolved puzzle of neutrino oscillations in core-collapse supernovae and neutron star mergers. Our analysis additionally reveals the previously unexplored phenomenon of flavor-wave synchronization.
\end{abstract}

\maketitle

\textbf{\textit{Introduction.}}---Outcomes of neutrino flavor mixing are uncertain in core-collapse supernovae and neutron star mergers \cite{volpe2024neutrinos, johns2025neutrino}. Although the equations that need to be solved are known, they apparently call for the resolution of impracticably small scales (see Refs.~\cite{shalgar2023neutrino, shalgar2024length, liu2025resolution} for debate concerning this point). A promising strategy is to implement some kind of coarse-grained treatment of neutrino transport \cite{li2021neutrino, zaizen2023simple, xiong2024robust, abbar2024physics, richers2024asymptotic, nagakura2024bhatnagar, johns2023thermodynamics, johns2024subgrid, johns2025local, johns2025implications, kost2024once, kost2025once, fiorillo2024fast, fiorillo2025collective, liu2024quasi, liu2025asymptotic}.

Recent work in this vein suggests that the effects of flavor mixing may be quite significant, with consequences for supernova explosions \cite{PhysRevLett.130.211401,ehring2023,PhysRevD.109.123008,10.1093/pasj/psaf007,Wang_2025,10.1093/mnras/stac3763}, post-merger disk evolution \cite{george2020fast,li2021,just2022,fernandez2022,PhysRevD.108.103014,h2q7-kn3v,lund2025angle}, nucleosynthesis and kilonovae, and neutrino and gravitational-wave emission \cite{sawyer2016neutrino, wu2015effects, wu2017fast, 
wu2017imprints, abbar2019occurrence, delfan2019linear, morinaga2020fast, xiong2020potential, ko2020neutrino, george2020fast, abbar2021characteristics, nagakura2021where, tamborra2021new, johns2021collisional, li2021, just2022, fernandez2022, fujimoto2023explosive, xiong2023collisional, liu2023universality, ehring2023, ehring2023fast, ehring2024gravitational, froustey2024neutrino, akaho2024collisional, liu2024muon, mukhopadhyay2024time, h2q7-kn3v, wang2025, lund2025angle, mori2025}. Most studies have concentrated on the effects of fast flavor conversion (FFC). It is possible that estimated effects will be further amplified by slow \cite{kostelecky1993neutrino, duan2006collective, duan2010collective, dasgupta2015temporal, chakraborty2016collective, shalgar2024neutrino, fiorillo2025theoryslow, fiorillo2025theoryslow2, fiorillo2025first} and collisional \cite{johns2021collisional, johns2022collisional, xiong2022evolution, xiong2022collisional, liu2023systematic, liu2023universality, akaho2023collisional, shalgar2023neutrinos, kato2023collisional, froustey2024neutrino, zaizen2025spectral, froustey2025predicting, z3qh-nj18} flavor instabilities.

Subgrid models of FFC are largely based on some notion of collisionless flavor equilibrium. Here we make progress on the theory of this centrally important concept by analyzing its characteristics across a range of dynamical regimes. In the course of our analysis, we identify neutrino analogues of some venerable phenomena in plasmas and self-gravitating systems governed by the Vlasov (or collisionless Boltzmann) equation.

Various features of Vlasov dynamics have previously been discussed in the neutrino context, including dispersion relations supporting absolute and convective instabilities \cite{banerjee2011linearized, sawyer2016neutrino, chakraborty2016self, izaguirre2017fast, capozzi2017fast, yi2019dispersion, morinaga2022fast, dasgupta2022collective, fiorillo2025theoryslow2, fiorillo2025dispersion, dasgupta2025sufficient} and  collisionless relaxation proceeding via phase-space filamentation, phase-space turbulence, and Landau damping \cite{sawyer2005speed, raffelt2007self, mangano2014damping, capozzi2019fast, johns2020fast, bhattacharyya2021fast, richers2022code, fiorillo2023slow, urquilla2024chaos, fiorillo2024theory}. These studies have clarified the mechanism by which a collisionless neutrino system relaxes to an asymptotic quasisteady state. However, they have not addressed the nature of the resulting flavor equilibrium.

In this paper we follow a path that mirrors the historical one taken in plasma physics. Early foundational work by Landau and others elucidated the  collisionless relaxation mechanism now known as Landau damping, in which plasma waves are damped by the loss of resonant particles from the collective motion. Analyses based on the linearized Vlasov equation left open the question of \textit{nonlinear} Landau damping: How does a collisionless plasma evolve when the waves are not small? O'Neil took a pivotal step by demonstrating that finite-amplitude waves, by trapping particles, can terminate or even reverse their own damping \cite{oneil1965collisionless, davidson1972methods}. For a system involving a single plasma wave, O'Neil showed the dynamics to be equivalent to that of a pendulum. Later, in another classic paper, Bernstein, Green, and Kruskal formulated the general notion of collisionless plasma equilibrium, which has come to be known as a BGK mode: a time-dependent solution of the nonlinear Vlasov--Poisson system, wherein the configuration is mutually determined by the particles and the waves \cite{bernstein1957exact, danielson2004measurement, ng2005bernstein}.\footnote{Despite the shared acronym, BGK modes are distinct from the Bhatnagar--Gross--Krook approximation \cite{bhatnagar1954model}, another concept recently adapted from plasmas to neutrinos \cite{nagakura2024bhatnagar}.} Numerical simulations reveal asymptotic states consisting of (superpositions of) BGK modes \cite{roberts1967nonlinear, galeotti2005asymptotic, carril2023formation}.

Analogues of these nonlinear particle--wave phenomena are already known in gravitational kinetics \cite{hamilton2024, lyndenbell1962stability, mathur1990existence, kandrup1998violent, kandrup1998collisionless, vandervoort2003, binney2008galactic, ng2021landau, tremaine2023dynamics}. In this work, in the context of a two-beam model of FFC, we explore the neutrino analogues. We show that single-wave FFC is equivalent to the motion of a flavor-wave pendulum, which is distinct from other neutrino pendulum solutions \cite{hannestad2006self, duan2007analysis, johns2018strange, johns2020neutrino, padilla2022neutrino, fiorillo2023slow, xiong2023symmetry, johns2023collisional, fiorillo2025fast} and was anticipated by Duan and collaborators~\cite{duan2009symmetries, duan2021flavor}. The stable pendulum fixed points are neutrino BGK modes, and the phase-space orbits alternate between wave damping and growth.  We then demonstrate the transition from this mechanical regime to an effectively irreversible one as the number of participating flavor waves is increased. The many-wave system collisionlessly relaxes to a dynamical equilibrium. We find evidence that a new phenomenon, flavor-wave synchronization, plays a role in dynamical equilibrium across all regimes and in equilibration of the many-wave system. Our findings provide new fundamental insights into neutrino flavor dynamics that will be useful in the effort to improve the accuracy of neutrino subgrid models.

\textbf{\textit{The two-beam model.}}---One of the simplest systems manifesting FFC consists of two head-on colliding neutrino beams in a 1D periodic spatial domain, $r\in[0,L]$. The evolution is described by (in units of $\mu = 1$)
\begin{align}   (\partial_t+\partial_r)\bm{P}_R &= 2\bm{P}_L\times\bm{P}_R, \notag \\
(\partial_t-\partial_r)\bm{P}_L &= 2\bm{P}_R\times\bm{P}_L. \label{eq:mainEOMs}
\end{align}
In the above equation, $\bm{P}_R$ and $\bm{P}_L$ are the polarization vectors of right- and left-moving neutrinos, respectively. The polarization vectors are initialized along the $z$-axis in flavor space, with small transverse perturbations to seed instability.

We use the Fourier decomposition $P^a_I (r)=\sum_{n\in\mathbb{Z}}P^a_{I,n}e^{i k_n r}$, where $a$ identifies the flavor-space component ($x, y, z$), $I$ identifies the beam ($R, L$), and $k_n \equiv 2\pi n/L$. The equations of motion in Fourier space are
\begin{align}
\dot{\bm{P}}_{R,n}=&-i k_n\bm{P}_{R,n}+2\sum_{m\in\mathbb{Z}}\bm{P}_{L,m-n}\times\bm{P}_{R,m}, \notag \\
\dot{\bm{P}}_{L,n}=&+i k_n\bm{P}_{L,n}+2\sum_{m\in\mathbb{Z}}\bm{P}_{R,m-n}\times\bm{P}_{L,m}. \label{eq:QKE_Fourier0}
\end{align}
Since $\bm{P}_I(r)$ is real, $\bm{P}_{I,n}$ obeys the complex conjugate relation $\bm{P}_{I,n}^*=\bm{P}_{I,-n}$.

To enhance the analytic tractability of the model, we impose $\bm{\hat{P}}_{I,0} = \pm \bm{z}$ and $\bm{z} \cdot \bm{P}_{I,n \neq 0} = 0$, which are good approximations in our later numerical results. We define spatially averaged sum and difference vectors $\bm{S}\equiv\bm{P}_{R,0}+\bm{P}_{L,0}$ and $ \bm{D}\equiv\bm{P}_{R,0}-\bm{P}_{L,0}$.
From Eqs.~\eqref{eq:QKE_Fourier0} it is clear that $\bm{S}$ is an invariant and
\begin{equation}
\dot{D}^z = 4 ( \bm{P}_{L,0} \times \bm{P}_{R,0} )^z + 8 \sum_{n>0} \mathrm{Re}\left[\bm{P}_{L,n}^*\times\bm{P}_{R,n}\right]^z.
\label{eq:EOM_D}
\end{equation}
The first term on the right vanishes because $\bm{\hat{P}}_{I,0} = \pm \bm{z}$.

\textbf{\textit{Staticity \& synchronization.}}---Our primary interest is in dynamical equilibria. Before coming to the physical parallels with other systems, we first consider from a mathematical standpoint what conditions must be met for the spatially averaged flavor configuration to be static.

First, the $n = 0$ polarization vectors must be oriented in such a way that they are constant under one another's influence, a notion that has been termed \textit{mixing equilibrium} \cite{johns2023thermodynamics, johns2024subgrid, johns2025local}. Mathematically, this first condition is $\bm{H}_{I,0} \times \bm{P}_{I,0} = 0$.
In this case, $\bm{H}_{R,0} = 2 \bm{P}_{L,0}$ and $\bm{H}_{L,0} = 2 \bm{P}_{R,0}$ [Eqs.~\eqref{eq:mainEOMs}]. The condition of mixing equilibrium is always satisfied here by virtue of $\bm{\hat{P}}_{I,0} = \pm \bm{z}$, but Eq.~\eqref{eq:EOM_D} makes clear that the spatially averaged system may still undergo evolution due to the coupling to the flavor waves. The second staticity condition is thus that the final term in Eq.~\eqref{eq:EOM_D} vanishes as well. Dynamical equilibrium is ensured when both conditions are met.

\begin{figure}
    \centering
    \includegraphics[width=.7\linewidth]{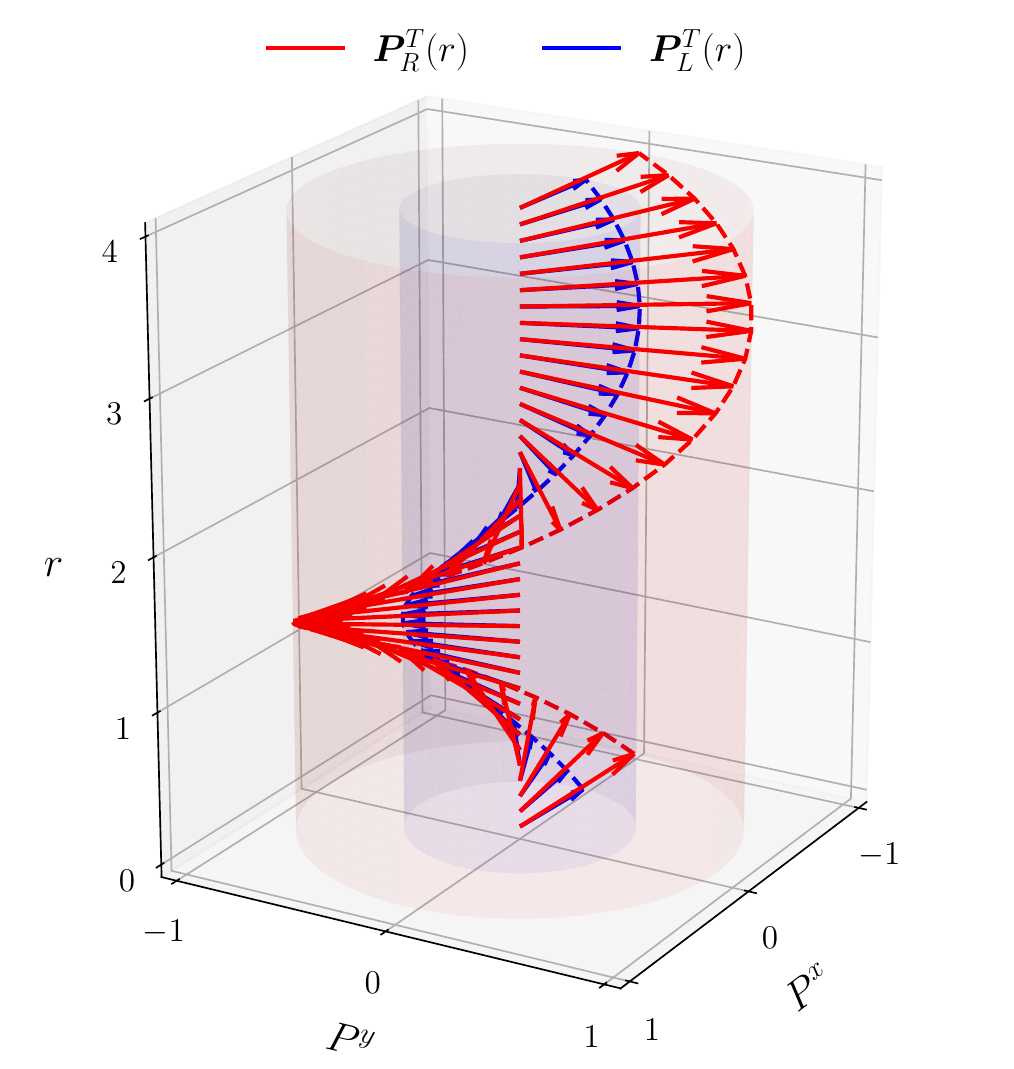}
    \caption{Illustration of phase-aligned, circularly polarized flavor-wave synchronization with a single inhomogeneous Fourier mode ($n = 1$). All $\bm{P}_I(r)$ co-precess around $\bm{z}$. This configuration is consistent with equilibrium spatial averages $\bm{P}_{I,0}$. By contrast, sustained phase misalignment causes secular change in $\bm{P}_{I,0}$.}
    \label{fig:cartoon}
\end{figure}

If $\bm{P}_{I,n}$ is an essentially random fluctuation, then $D^z$ is approximately constant despite the coupling \cite{johns2023thermodynamics, johns2024subgrid, johns2025local}. In this sense dynamical equilibrium can be supported by subgrid ($n \neq 0$) disorder. However, this is not the only possibility. Staticity is also consistent with highly ordered flavor waves. In particular, consider the following configuration. Write $P^{a}_{I,n} = \left|P^{a}_{I,n}\right|e^{i\varphi^{a}_{I,n}}$ for $a = x, y$ and define the relative phases
\begin{equation}
    \delta_n^{a} \equiv \varphi_{R,n}^{a} - \varphi_{L,n}^{a}, ~~~ \vartheta_{I,n} \equiv \varphi^y_{I,n} - \varphi^x_{I,n}.
\end{equation}
Take the flavor waves to be phase-aligned ($\delta_n^y = \delta_n^x = 0, \pm \pi$) and circularly polarized ($\vartheta_{R,m} = \vartheta_{L,m} = \pm \pi/2$). It is easy to confirm using Eq.~\eqref{eq:EOM_D} that $D^z$ is constant under these circumstances (see Sec.~I of the Supplemental Material). If phase alignment is relaxed ($\delta_n^x \neq 0$) but synchronization still holds ($\dot{\delta}_n^x = 0$), then $D^z$ changes secularly.

Phase-aligned synchronization is visualized in Fig.~\ref{fig:cartoon} for the single mode $n = 1$. It is essentially a spatially inhomogeneous generalization of the concept of pure precession \cite{raffelt2007self2, duan2009symmetries}, although pointwise alignment of $\bm{P}_R$ and $\bm{P}_L$ as functions of $r$ generally does not occur with more than one mode even if $\delta^x_n = 0$ for all $n$. While the concept of flavor-wave synchronization is special to neutrinos, we will return to it repeatedly in developing the parallels with other mean-field kinetic systems.

\begin{figure*}
    \centering
    \includegraphics[width=1\linewidth]{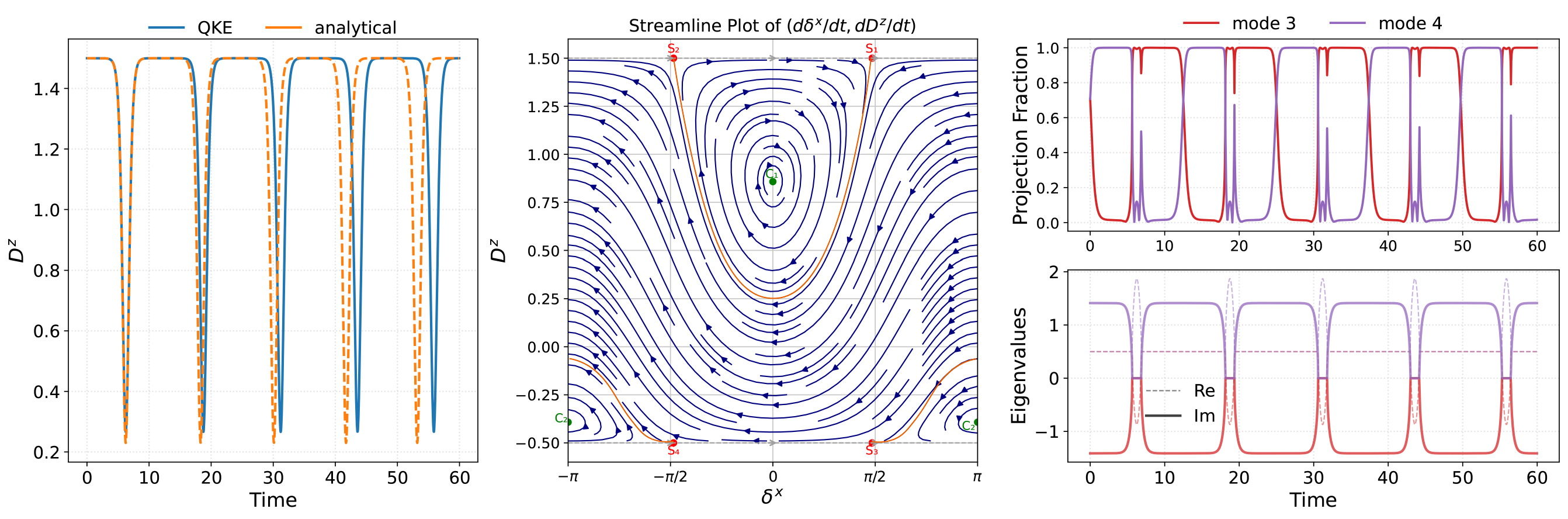}
    \caption{Comparison between the full QKE simulation and the reduced model for mode $n=1$ in a box of size $L=4$. Left: Periodic fast flavor oscillations using identical initial conditions with single-mode transverse seed perturbations. Middle: Phase portrait showing completely regular dynamics. Right: Eigenmode decomposition of the QKE solution, with projection coefficients normalized to unit total power.}

    \label{fig:regular}
\end{figure*}

\textbf{\textit{Regular motion: A single wave.}}---We now seek neutrino flavor dynamics analogous to O'Neil's pendulum solution for the motion of electrons in a single-wave plasma \cite{oneil1965collisionless, davidson1972methods}. Based on prior literature, it is not clear that a single flavor wave can be isolated in practice in the nonlinear regime. We have discovered a technique for doing so: the number of significantly active Fourier modes can be tuned by adjusting the size of the box $L$. This is possible because fast instabilities occur only within a finite wavenumber interval $(k_{\mathrm{min}},k_{\mathrm{max}})$. While this range is independent of the box size, the wave vectors that are realized in the system are quantized in units of $2 \pi /L$. For this part of the analysis, we choose $L$ such that only a single mode, $n = 1$, falls within the unstable interval.

The exact numerical solution in the left panel of Fig.~\ref{fig:regular} confirms, surprisingly, that spatially inhomogeneous FFC is indeed capable of periodic motion. We can solve for the evolution semi-analytically by reducing the system to two relevant dynamical degrees of freedom, $D^z$ and $\delta^x \equiv \delta_1^x$, which evolve under
\begin{align}
    \dot{D}^z &= - 4 \mathcal{F}_R \mathcal{F}_L \sin \delta^x, \notag\\
    \dot{\delta}^x &= -2 k + 2 D^z + [ \mathcal{F}_R (S^z + D^z) / \mathcal{F}_L \notag \\
    &\hspace{1.05 in} - \mathcal{F}_L (S^z - D^z) / \mathcal{F}_R ] \cos \delta^x, \label{eq:fw_pend}
\end{align}
where $k \equiv k_1$ and $\mathcal{F}_{R(L)} \equiv (|\bm{P}_{R,L}|^2 - (S^z \pm D^z)/4)^{1/2}$. The derivation of these equations is presented in Sec.~II of the Supplemental Material. The left panel of Fig.~\ref{fig:regular} compares the solution of Eqs.~\eqref{eq:fw_pend} to the exact numerical solution for the representative case with $L = 4$. The solutions agree quantitatively apart from a growing discrepancy due to the spurious appearance of flavor waves in other Fourier modes generated by numerical diffusion in the finite volume QKE solver.

The middle panel shows flow lines in the $(D^z, \delta^x)$ phase space, exhibiting librating orbits surrounding the stable fixed points (the centers $C_1$ and $C_2$), heteroclinic orbits connecting unstable fixed points (the saddle-point pairs $S_1$--$S_2$ and $S_3$--$S_4$), and circulating orbits outside of the trapping regions. The flow direction on the boundaries ($D^z=1.5$ and $D^z=-0.5$) is only defined in the limit. The evolution in the left panel corresponds to periodic oscillation between $S_1$ and $S_2$. Depending on the initial seed, the oscillation can be either librating or circulating. Notice that the stable fixed points $C_1$ and $C_2$ are marked by phase-aligned flavor wave synchronization ($\delta^x = 0, \pm \pi$). Their locations can be found by solving Eqs.~\eqref{eq:fw_pend} with $\dot{D}^z = \dot{\delta}^x = 0$.

We gain further insight into the flavor-wave pendulum using eigenanalysis. Following Ref.~\cite{johns2025local}, we recast the flavor waves in terms of $\Psi_n \equiv ( \bm{P}_{R,n}, \bm{P}_{L,n})^T$ and linearize Eqs.~\eqref{eq:QKE_Fourier0} in $\Psi_n$ to obtain a Schr\"{o}dinger-like equation of motion $i \partial_t \Psi_n = H_n \Psi_n$ with non-Hermitian Hamiltonian $H_n$. (Calculational details, including explicit expressions for the eigenmodes $\psi$ and eigenvalues $\Omega$, are found in Sec.~III of the Supplemental Material.) In the right panels of Fig.~\ref{fig:regular}, the QKE solution for $\Psi_1$ is decomposed into the instantaneous eigenbasis of $H_1$. The flavor wave alternates between two eigenmodes, repeatedly passing through exceptional points where the eigenmodes become degenerate. Only the pair of complex eigenmodes contributes to the full dynamics, whereas the other oscillatory modes remain negligible if they are initially seeded perturbatively. $D^z$ alternately dips and rebounds as $\Psi_1$ grows and decays. We have confirmed that the stable fixed point decomposes into a single eigenmode with $\textrm{Im}(\Omega) = 0$.

The early growth of the initially unstable flavor wave (labeled mode 4 following the numbering in the Supplemental Material) is an instance of inverse linear Landau damping \cite{fiorillo2024theory}. Eventually the growth feeds back significantly on $D^z$, which in turn alters the flavor-wave properties. This bidirectional influence defines the regime of nonlinear Landau damping and makes it possible for damping to saturate. While saturation does occur in the many-wave system (see below), nonlinear Landau damping in the single-wave system involves alternating phases of energy transfer from $D^z$ to $\Psi_1$ and back again.

The stable fixed points of the flavor-wave pendulum are the simplest example of collisionless equilibrium supported by flavor-wave synchronization. They are neutrino BGK modes, being static solutions of the full nonlinear equations. We next examine how this picture generalizes in the presence of multiple waves.

\begin{figure}
    \centering
    \includegraphics[width=0.95\linewidth]{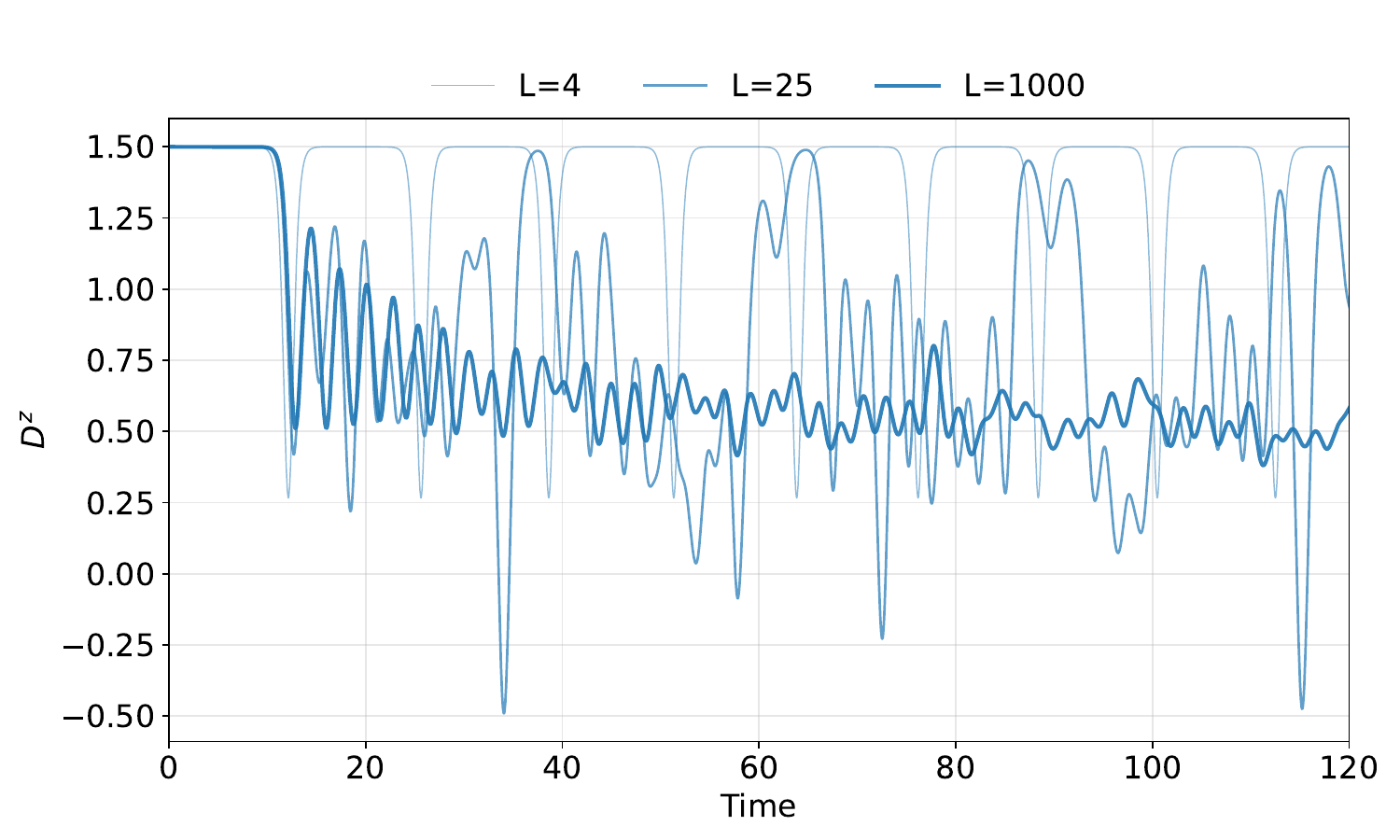}
    \caption{Time evolution of $D^z$ for box sizes $L = 4$, 25, and 1000, corresponding to 1, 11, and 450 unstable Fourier modes, respectively. As the number of unstable modes increases, the system transitions from coherent recurrences to chaotic evolution and effectively irreversible relaxation.}

    \label{fig3}
\end{figure}

\textbf{\textit{Effective irreversibility: Many waves.}}---Our strategy of varying $L$ grants us the ability to show how FFC transitions from regular to effectively irreversible dynamics as the number of active Fourier modes approaches the continuum limit. We vary the box size  using the same initial setup but with random multi-mode transverse perturbations. As $L$ increases, more modes fall within the unstable range. The $D^z$ evolution is plotted in Fig.~\ref{fig3}. Intermode dephasing suppresses recurrence in systems with many flavor waves, which interact directly through the final term in Eq.~\eqref{eq:EOM_D} and indirectly through fluctuations in $D^z$. This parallels the nonlinear Landau damping in the multi-wave plasma setting, where overlapping resonances and dephasing lead to irreversibility and drive the system toward a quasi-stationary state.

\begin{figure}
    \centering
    \includegraphics[width=1\linewidth]{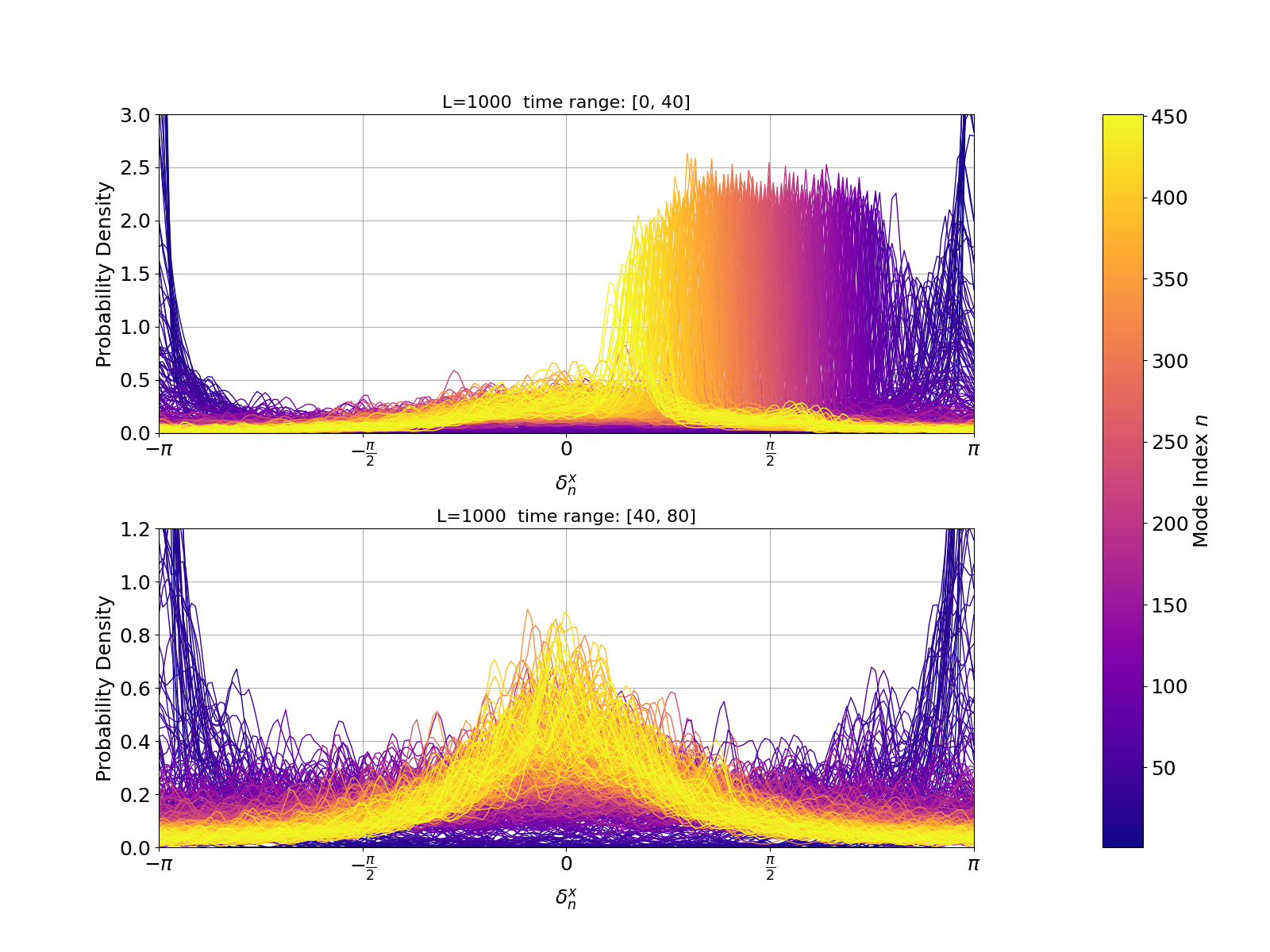}
    \caption{Probability density of the flavor-wave phase difference $\delta^x_n$ for each mode $n$ in the $L = 1000$ simulation. Upper panel: nonequilibrium regime ($t \in [0, 40]$) showing phase misalignment. Lower panel: equilibrium regime ($t \in [40, 80]$) showing phase alignment. Color indicates mode index $n$. The total probability for each mode $n$ is normalized to unity.}

    \label{fig4}
\end{figure}

Although the generalization of the pendulum phase space is not obvious, we can generalize flavor-wave alignment by defining a probability distribution function $f(\delta^x_n)$ for each $n$, with the ensemble constructed by sampling points over some time window. Inspection of $f$ reveals whether flavor-wave alignment occurs in a time-averaged sense. In Fig.~\ref{fig4} we present this distribution for modes up to the maximum unstable mode. Distributions are sampled in two time windows: the out-of-equilibrium period $t \in [0, 40]$ (upper panel) and the in-equilibrium period $t \in [40, 80]$ (lower panel). The nonequilibrium dynamics shows broad distributions covering the region around $\delta_n^x \approx \pi/2$ and narrow distributions around $\pm \pi$. By contrast, the equilibrium dynamics is concentrated around $\delta_n^x = 0, \pm\pi$. 

In both panels the lowest 10 Fourier modes, which are all linearly stable, exhibit concentrated distributions near $\pm\pi$. For them, the pair $S_1S_2$ dominating the delay between oscillations is near $\pm\pi$, leading to the high distribution density in that region. On the other hand, the distributions determine the staticity of the coarse-grained dynamics. Phase-misaligned synchronization, characterized by peaks around $\pi/2$, drives secular evolution of $D^z$, whereas phase-aligned synchronization, with peaks around $0$, permits staticity by making the summation in Eqs.~\eqref{eq:QKE_Fourier0} vanish. This connection is formalized in Sec.~I of the Supplemental Material. The equilibrium distributions in Fig.~\ref{fig4} (bottom) therefore contrast with the circulating streamlines of Fig.~\ref{fig:regular} (middle), which illustrate periodic precession in the single-wave pendulum regime.

Suppose that we approximate Eqs.~\eqref{eq:QKE_Fourier0} using the quasilinear approximation \cite{fiorillo2024fast}. Then, in the case that there is a single eigenmode at each $n$, with $\textrm{Im}(\Omega) = 0$ for every one of these modes, $D^z$ is exactly constant (see Sec.~III of the Supplemental Material for the proof). Such a state is a quasilinear equilibrium whose staticity is supported by phase-aligned flavor-wave synchronization. As noted earlier, equilibrium can also be supported by uncorrelated subgrid fluctuations \cite{johns2023thermodynamics, johns2024subgrid, johns2025local}. Figure~\ref{fig4} is evidence that in reality both mechanisms are likely to play a role, with the relative importance of disorder here reflected in the broadening of $f$ around the $\delta^x_n = 0, \pm \pi$ peaks.

\textbf{\textit{Discussion.}}---Collisionless equilibria are of central importance in physical systems governed by mean-field interactions. In this work we have developed fundamental insights into collisionless equilibria in dense neutrino systems. Such equilibria are the cornerstone of most approaches currently being developed for the incorporation of neutrino oscillations into simulations of core-collapse supernovae and neutron star mergers. We have also drawn parallels with well-established phenomena in plasmas and self-gravitating systems: nonlinear Landau damping, the single-wave pendulum, trapped and passing orbits, and collisionless equilibria.

Running throughout our analysis are the new concepts of flavor-wave synchronization, alignment, and polarization. The phase synchronization we identify calls to mind---but is distinct from---the type of synchronization that motivated Pantaleone to stress the parallels between neutrino systems and the Kuramoto model of phase oscillators \cite{pantaleone1998stability}. The significance of phase synchronization in kinetic plasmas is also increasingly appreciated, where it is understood that synchronization competes with filamentation to limit entropy growth \cite{xu2021, ghizzo2022, ghizzo2023}.

Deviation from phase alignment feeds into the response of neutrino systems to astrophysical driving. This may be a useful insight for coarse-grained theories designed to approximate exactly that response \cite{johns2023thermodynamics, johns2024subgrid, johns2025local, fiorillo2024fast, fiorillo2025collective}. On the other hand, we have also called attention to an issue that these theories most likely need to address: exceptional points, which coincide with marginal stability, threaten the approximation that flavor waves track eigenmodes without transitions due to finite rates of change. (This approximation is called \textit{quasistatic} in Ref.~\cite{johns2025local} and \textit{WKB} in Ref.~\cite{fiorillo2024fast}.) Exceptional points are evident in Fig.~\ref{fig:regular} and are also observed in our large-$L$ calculations. We leave for future work the resolution of this issue and the further leveraging of flavor-wave dynamical characteristics to improve existing coarse-grained transport proposals.

\textbf{\textit{Acknowledgments.}}---
LJ is supported by a Feynman Fellowship through LANL LDRD project number 20230788PRD1.
M.Z. is supported by JSPS KAKENHI Grant Nos. JP24H02245 and JP25K17383. 
HN is supported by Grant-in-Aid for Scientific Research (23K03468), the NINS International Research Exchange Support Program, and the HPCI System Research Project (Project ID: hp250006, hp250226, hp250166).
S.Y. is supported by Grant-in-Aid for Scientific Research (25K01006) and the Waseda University Grant for Special Research Projects (project No. 2025C-136). He is also supported by the Institute for Advanced Theoretical and Experimental Physics, Waseda University.

\bibliographystyle{apsrev4-2}
\bibliography{refs}

\onecolumngrid
\clearpage

\setcounter{equation}{0}\setcounter{figure}{0}\setcounter{table}{0}
\renewcommand{\theequation}{S\arabic{equation}}
\renewcommand{\thefigure}{S\arabic{figure}}
\renewcommand{\thetable}{S\arabic{table}}

\makeatletter
\newcounter{smsec}
\renewcommand{\thesmsec}{\Roman{smsec}}     
\providecommand{\theHsmsec}{\Roman{smsec}}  
\newcommand{\SMsection}[2][]{%
  \refstepcounter{smsec}
  \section*{\thesmsec.\ #2}%
  \ifx\relax#1\relax\else\label{#1}\fi
}
\makeatother


\begin{center}
\textbf{\large Supplemental Material for\\
\emph{Dynamical equilibria of fast neutrino flavor conversion}}\\[6pt]
Jiabao Liu$^{1}$, Lucas Johns$^{2}$, Hiroki Nagakura$^{3}$, Masamichi Zaizen$^{4}$, Shoichi Yamada$^{1,5}$\\[3pt]
\textit{$^{1}$Department of Physics and Applied Physics, School of Advanced Science \& Engineering, Waseda University, Tokyo 169-8555, Japan}\\
\textit{$^{2}$Theoretical Division, Los Alamos National Laboratory, Los Alamos, NM 87545, USA}\\
\textit{$^{3}$Division of Science, National Astronomical Observatory of Japan, 2-21-1 Osawa, Mitaka, Tokyo 181-8588, Japan}\\
\textit{$^{4}$Department of Earth Science and Astronomy, The University of Tokyo, Tokyo 153-8902, Japan}\\
\textit{$^{5}$Research Institute for Science and Engineering, Waseda University, Tokyo 169-8555, Japan}
\end{center}

We provide technical details that support the findings reported in the main text. Specifically, we (i) relate dynamical equilibrium to flavor-wave synchronization (Sec.~\ref{sm:synch}), (ii) derive the flavor-wave pendulum equations of motion (Sec.~\ref{sm:pend}), (\ref{sm:eigen}) present the flavor-wave eigenanalysis (Sec.~\ref{sm:eigen}), and (iv) discuss the classification of collisionless neutrino equilibria (Sec.~\ref{sm:equil}).


\refstepcounter{section}
\SMsection[sm:synch]{Flavor-wave synchronization}

To begin, we recap the definitions of some of the relevant variables. We define
\begin{equation}
D^z \equiv P^z_{R,0}-P^z_{L,0}, \qquad
S^z \equiv P^z_{R,0}+P^z_{L,0},
\end{equation}
where $P^z_{I,0}$ is the $z$–component of the $I$th beam's spatially averaged polarization vector. Recall that the transverse parts $\bm{P}^T_{I,0}$ are assumed to vanish. Meanwhile, the flavor waves (the $n \neq 0$ Fourier modes) $\bm{P}_{I,n}$ are assumed to be purely transverse: $P^z_{I,n} = 0$.

Since $S^z$ is constant and the transverse parts vanish, the spatially coarse-grained dynamics is entirely captured by the equation of motion
\begin{equation}
    \dot{D}^z = 4 (\bm{P}_{L,0} \times \bm{P}_{R,0})^z + 8 \sum_{n>0} \mathrm{Re}\left[\bm{P}_{L,n}^*\times\bm{P}_{R,n}\right]^z.
\end{equation}
This follows straightforwardly from the quantum kinetic equations in Fourier space after eliminating all negative Fourier modes using the conjugate relation $P^a_{I,-n}=(P^a_{I,n})^*$. We use \textit{dynamical equilibrium} to refer to a collisionless system that displays no coarse-grained evolution. In this case, a system is in dynamical equilibrium if and only if
\begin{equation}
    D^z = \textrm{constant}.
\end{equation}
A state of \textit{mixing equilibrium} is one for which the spatially averaged polarization vectors are in a mutually static configuration, without regard for the influence of the flavor waves. The condition of mixing equilibrium here is
\begin{equation}
  \bm{P}_{L,0} \times \bm{P}_{R,0} = 0, 
\end{equation}
which is satisfied by virtue of the assumption $\bm{P}^T_{L,0} = \bm{P}^T_{R,0} = 0$. Therefore
\begin{equation}
    \dot{D}^z = 8 \sum_{n>0} \mathrm{Re}\left[\bm{P}_{L,n}^*\times\bm{P}_{R,n}\right]^z.
\end{equation}
The system is always in mixing equilibrium, but it may nevertheless undergo evolution due to the coupling of the flavor waves to $D^z$. The condition
\begin{equation}
    \sum_{n > 0} \mathrm{Re}\left[\bm{P}_{L,n}^*\times\bm{P}_{R,n}\right]^z = 0
\end{equation}
must be satisfied to ensure that the system is in dynamical equilibrium.

Let us decompose the components of $\bm{P}_{I,n}$ into their norms and phases:
\begin{equation}
    P^a_{I,n} = | P^a_{I,n} | e^{i \varphi^a_{I,n}}. \label{eq:defPnorm}
\end{equation}
Consider a dynamical equilibrium for which
\begin{equation}
    \mathrm{Re}\left[\bm{P}_{L,n}^*\times\bm{P}_{R,n}\right]^z = 0 \label{eq:strictdyneq}
\end{equation}
for each $n$ individually. Expanding out the left-hand side of Eq.~\eqref{eq:strictdyneq} using Eq.~\eqref{eq:defPnorm},
\begin{equation}
    \mathrm{Re}\left[\bm{P}_{L,n}^*\times\bm{P}_{R,n}\right]^z = | P_{L,n}^x | |P_{R,n}^y | \cos (- \varphi_{L,n}^x + \varphi_{R,n}^y) - | P_{L,n}^y | |P_{R,n}^x | \cos (- \varphi_{L,n}^y + \varphi_{R,n}^x). \label{eq:Revarphi}
\end{equation}
Now we introduce the relative phases
\begin{equation}
    \delta_n^{a} \equiv \varphi_{R,n}^{a} - \varphi_{L,n}^{a}, \qquad \vartheta_{I,n} \equiv \varphi^y_{I,n} - \varphi^x_{I,n}
\end{equation}
and rewrite Eq.~\eqref{eq:Revarphi} as
\begin{equation}
    \mathrm{Re}\left[\bm{P}_{L,n}^*\times\bm{P}_{R,n}\right]^z = | P_{L,n}^x | |P_{R,n}^y | \cos (\delta_n^x + \vartheta_{R,n}) - | P_{L,n}^y | |P_{R,n}^x | \cos (\delta_n^x - \vartheta_{L,n}).
\end{equation}
Observe that if all flavor waves at a given $n$ have uniformly right-handed (RH) or left-handed (LH) polarization,
\begin{equation}
    \vartheta_{R,n} = \vartheta_{L,n} = \pm \frac{\pi}{2}, \label{eq:RHLHpol}
\end{equation}
then 
\begin{equation}
    \mathrm{Re}\left[\bm{P}_{L,n}^*\times\bm{P}_{R,n}\right]^z = \mp \left( | P_{L,n}^x | |P_{R,n}^y | + | P_{L,n}^y | |P_{R,n}^x | \right) \sin \delta_n^x. \label{eq:Rezmp}
\end{equation}
Hence the influence of flavor waves on the coarse-grained system vanishes (\textit{i.e.}, $D^z = \textrm{constant}$) for
\begin{equation}
    \delta_n^x = 0, \pm \pi. \label{eq:phasealigned}
\end{equation}
This proves the claim in the main text that this type of configuration---phase-aligned [Eq.~\eqref{eq:phasealigned}] flavor waves with uniform RH or LH polarization [Eq.~\eqref{eq:RHLHpol}]---exerts no change in $D^z$. Flavor-wave synchronization
\begin{equation}
    \dot{\delta}_n^x = 0 \label{eq:FWsynchx}
\end{equation}
in one of these uniform polarization configurations guarantees that the system is in dynamical equilibrium. Such motion is similar to the pure-precession solutions of Refs.~\cite{raffelt2007self2, duan2009symmetries}. We will see in Sec.~\ref{sm:eigen} that if a single stable eigenmode is occupied at a given $n$, then $\bm{P}_{R,n}$ and $\bm{P}_{L,n}$ do indeed undergo precession motion satisfying Eqs.~\eqref{eq:RHLHpol} and \eqref{eq:phasealigned}.

A system may be in a fluctuating dynamical equilibrium such that $D^z$ is not strictly constant but its average over time window $\tau$ is:
\begin{equation}
    \overline{D^z} (t) \equiv \frac{1}{\tau} \int_{t-\tau}^t dt' D^z (t') =  \textrm{constant}. \label{eq:Dzoverline}
\end{equation}
This is the possibility identified in Refs.~\cite{johns2023thermodynamics, johns2024subgrid}, with the influence of flavor waves on the mean vanishing due to the $\bm{P}_{I,n}$ motion being effectively random. In phase-aligned dynamical equilibrium, staticity is maintained by order rather than disorder in the flavor waves. We will see in Sec.~\ref{sm:eigen} that if multiple stable eigenmodes are occupied at a given $n$ (and no unstable eigenmodes), then randomization in the sense above [Eq.~\eqref{eq:Dzoverline}] is likely.

As another contrast with phase-aligned dynamical equilibrium, consider a system with flavor-wave synchronization [Eq.~\eqref{eq:FWsynchx}] in a phase-\textit{misaligned} configuration:
\begin{equation}
    \delta_n^x \neq 0, \pm \pi.
\end{equation}
By Eq.~\eqref{eq:Rezmp} flavor waves at this wave number promote a secular change in $D^z$. We will see in Sec.~\ref{sm:eigen} that flavor-wave instability coincides with phase-misaligned synchronization with uniform RH or LH polarization. Thus these three regimes---dynamical equilibrium supported by phase-aligned synchronization, dynamical equilibrium supported by flavor-wave randomization, and a secularly changing mixing equilibrium supported by phase-misaligned synchronization---reflect the eigenmode structures and overlaps.

\SMsection[sm:pend]{The flavor-wave pendulum}

To illustrate the above results, we specialize to a system with a single unstable Fourier mode. We suppress the Fourier mode subscript in this section and specify the single mode with wavenumber $k$. We restrict to initial conditions where the initial background flavor fields are homogeneous. We take the flavor waves to have RH polarization:
\begin{equation}
    \vartheta_I=\varphi^y_{I}-\varphi^x_{I} = +\pi/2. \label{eq:pendRH}
\end{equation}
In a flavor-wave system with a single polarization, we have 
\begin{equation}
    \delta^y=\varphi^y_R-\varphi^y_L=(\varphi^x_R+\pi/2)-(\varphi^x_L+\pi/2)=\delta^x.
    \label{eq:delta_x=y}
\end{equation}
This setup offers a neat relation between the norm of flavor waves
\begin{equation}
    |P^x_{I}|=|P^y_{I}|=|\bm{P}^T_{I}(r)|/2.
    \label{eq:FWnorm}
\end{equation}
The amplitude of the transverse polarization is homogeneous in space and is determined self-consistently from
\begin{equation}
    |\bm{P}^T_{I}| = \sqrt{|\bm{P}_{I}|^2 - (P^z_{I,0})^2}.
\end{equation}
For now we simply adopt this as an ansatz. The eigenanalysis of Sec.~\ref{sm:eigen} confirms that the single-wave system studied in the main text exhibits instability of eigenmodes with this structure, thus justifying Eq.~\eqref{eq:pendRH}.

This subspace is invariant---even under the full nonlinear equations of motion---in the sense that neither other Fourier modes nor opposite-polarization waves at the same mode are generated. The invariance follows from two points. First, the conditions of Eqs.~\eqref{eq:pendRH} and \eqref{eq:FWnorm} are preserved: taking time derivatives of $\vartheta_I$ and $|P^a_I|$, substituting the QKE, and reimposing these conditions cancels the derivative terms. Second, a RH flavor wave vector is proportional to $(1,i,0)^T$, so cross products of such vectors vanish, preventing the generation of additional Fourier modes.

Within this subspace, we then derive the pendulum-like equations of motion quoted in the main text:
\begin{equation}
\begin{split}
    \dot{D}^z &= -4\sqrt{\bigg[|\bm{P}_R|^2 - (S^z+D^z)^2/4\bigg]\bigg[|\bm{P}_L|^2-(S^z-D^z)^2/4\bigg]}\sin(\delta^x), \\
    \dot{\delta}^x &= -2k+2D^z +\Bigg(
    \sqrt{\tfrac{|\bm{P}_L|^2-(S^z-D^z)^2/4}{|\bm{P}_R|^2-(S^z+D^z)^2/4}}\,(S^z+D^z)-\sqrt{\tfrac{|\bm{P}_R|^2-(S^z+D^z)^2/4}{|\bm{P}_L|^2-(S^z-D^z)^2/4}}\,(S^z-D^z)\Bigg)\cos(\delta^x). \label{eq:pendEOM}
\end{split}
\end{equation}
They follow directly from the Fourier-space QKEs. The algebra is elementary but lengthy; we thus outline the main steps here: the equation for $D^z$ is obtained by inserting Eq.~\eqref{eq:Rezmp}, while the equation for $\delta^x$ arises from differentiating $\varphi^a_I=\arctan[\mathrm{Im}(P^a_I)/\mathrm{Re}(P^a_I)]$ and using Eqs.~\eqref{eq:pendRH}, \eqref{eq:FWnorm}, and \eqref{eq:delta_x=y} to eliminate redundant variables. Periodicity in this subspace holds strictly. Remarkably, the solutions to this single-wave pendulum are also \textit{exact} solutions to the full nonlinear QKE.

In this system $\sigma^x$ defined as
\begin{equation}
    \sigma^{a}\equiv\varphi^{a}_{R}+\varphi^{a}_{L}.
\end{equation}
is an independent dynamical variable in addition to $D^z$ and $\delta^x$, but its motion is completely determined by the other two. For completeness, its equation of motion is
\begin{equation}
\dot{\sigma}^x=-2S^z+\left(\sqrt{\frac{\left|\bm{P}_L\right|^2-(S^z-D^z)^2/4}{\left|\bm{P}_R\right|^2-(S^z+D^z)^2/4}}(S^z+D^z)+\sqrt{\frac{\left|\bm{P}_R\right|^2-(S^z+D^z)^2/4}{\left|\bm{P}_L\right|^2-(S^z-D^z)^2/4}}(S^z-D^z)\right)\cos(\delta^x).
\end{equation}

In the main text we confirm the validity of the analysis above by comparing the solution of Eqs.~\eqref{eq:pendEOM} to the solution of the full equations of motion.

\SMsection[sm:eigen]{Flavor-wave eigenanalysis}

The full equations of motion for $\bm{P}_{R,n \neq 0}$ and $\bm{P}_{L,n \neq 0}$ are
\begin{align}
\dot{\bm{P}}_{R,n}=&-i k_n\bm{P}_{R,n}+2\sum_{m\in\mathbb{Z}}\bm{P}_{L,m-n}\times\bm{P}_{R,m}, \notag \\
\dot{\bm{P}}_{L,n}=&+i k_n\bm{P}_{L,n}+2\sum_{m\in\mathbb{Z}}\bm{P}_{R,m-n}\times\bm{P}_{L,m}.
\label{eq:QKE_Fourier}
\end{align}
We adopt the assumption of weak inhomogeneity, in which $\bm{P}_{I,n'}$ for any $n' \neq 0$ is treated as small \cite{johns2025local} in the sense that it only contributes linearly to the source term of flavor waves. Note that this approximation does not assume that flavor coherence $\bm{P}_I^T(r)$ is small and is therefore applicable beyond the early stage of unstable evolution. Linearizing in $\bm{P}_{I,n'}$, we obtain
\begin{align}
\dot{\bm{P}}_{R,n}=&-i k_n\bm{P}_{R,n} + 2\bm{P}_{L,0}\times\bm{P}_{R,n} + 2\bm{P}_{L,n}\times\bm{P}_{R,0}, \notag \\
\dot{\bm{P}}_{L,n}=&+i k_n\bm{P}_{L,n} + 2\bm{P}_{R,0}\times\bm{P}_{L,n} + 2\bm{P}_{R,n}\times\bm{P}_{L,0}. \label{eq:linEOMs}
\end{align}
We now restrict our attention to flavor wave modes with positive wavenumbers
\begin{equation}
    k_n \equiv 2\pi n/L > 0
\end{equation}
because the $n < 0$ polarizations are related to the positive ones through conjugation: $\bm{P}_{I,-n} = \bm{P}_{I,n}^*$.

Equations~\eqref{eq:linEOMs} imply that the vector $\Psi_n \equiv (\bm{P}_{R,n} \ \ \bm{P}_{L,n})^T$ evolves under the Schr\"{o}dinger equation
\begin{equation}
    i \frac{d\Psi_n}{dt} = H_n \Psi_n
\end{equation}
with non-Hermitian Hamiltonian
\begin{equation}
    H_n =
    \begin{pmatrix}
        k_n & -i(S^z-D^z) & 0 & 0 & i(S^z+D^z) & 0 \\
        i(S^z-D^z) & k_n & 0 & -i(S^z+D^z) & 0 & 0 \\
        0 & 0 & k_n & 0 & 0 & 0 \\
        0 & i(S^z-D^z) & 0 & -k_n & -i(S^z+D^z) & 0 \\
        -i(S^z-D^z) & 0 & 0 & i(S^z+D^z) & -k_n & 0 \\
        0 & 0 & 0 & 0 & 0 & -k_n
    \end{pmatrix}.
\end{equation}
The six eigenvalues are
\begin{align}
    \Omega_{n,1(2)} &= \mp k_n, \\
    \Omega_{n,3(4)} &= +S^z \mp \sqrt{(S^z)^2+k_n^2-2D^z k_n}, \label{eq:Omega34}\\
    \Omega_{n,5(6)} &= -S^z \mp \sqrt{(S^z)^2+k_n^2+2D^z k_n}, \label{eq:Omega56}
\end{align}
where $\mp$ corresponds to the choice of index $i$ or $j$ in $\Omega_{n,i(j)}$. There can be at most one unstable eigenmode at each Fourier mode $n$ at a time. Instability occurs if and only if one of these eigenvalues acquires an imaginary part, which happens for
\begin{equation}
    |D^z| > D^z_{n,c} \equiv \frac{(S^z)^2 + k_n^2}{2k_n}.
\end{equation}

The first two eigenvectors are the noncollective modes 
\begin{equation}
    \psi_{n,1} = \begin{pmatrix}
        0 \\ 0 \\ 0 \\ 0 \\ 0 \\ 1
    \end{pmatrix}, \quad
    \psi_{n,2} = \begin{pmatrix}
        0 \\ 0 \\ 1 \\ 0 \\ 0 \\ 0
    \end{pmatrix}.
\end{equation}
The (un-normalized) collective eigenmodes are
\begin{equation}
    \psi_{n,3(4)} = 
    \begin{pmatrix}
        -i \zeta_{n,3(4)} \\ \zeta_{n,3(4)} \\ 0 \\ -i \\ 1 \\ 0
    \end{pmatrix}, \qquad
    \psi_{n,5(6)} = 
    \begin{pmatrix}
        i \zeta_{n,5(6)} \\ \zeta_{n,5(6)} \\ 0 \\ i \\ 1 \\ 0
    \end{pmatrix}, \label{eq:collectivemodes}
\end{equation}
where
\begin{align}
    \zeta_{n,3(4)} &= \frac{-D^z + k_n \mp \sqrt{(S^z)^2+k_n^2-2D^z k_n}}{D^z-S^z}, \\
    \zeta_{n,5(6)} &= \frac{-D^z - k_n \pm \sqrt{(S^z)^2+k_n^2+2D^z k_n}}{D^z-S^z}.
\end{align}
Note that
\begin{equation}
    \textrm{Im}(\zeta_{n,3}) = - \textrm{Im}(\zeta_{n,4}), \qquad
    \textrm{Im}(\zeta_{n,5}) = - \textrm{Im}(\zeta_{n,6}).
\end{equation}
When $\textrm{Im}(\Omega_{n,3}) \neq 0$, $\zeta_{n,3}^* = \zeta_{n,4}$. When $\textrm{Im}(\Omega_{n,3}) = 0$, $\zeta_{n,3}$ and $\zeta_{n,4}$ are both real. At the critical $D^z = D^z_{n,c}$, $\zeta_{n,3} = \zeta_{n,4}$ and so $\psi_{n,3} = \psi_{n,4}$. The eigenvectors coalesce at an exceptional point. The same holds for $\psi_{n,5}$ and $\psi_{n,6}$ if $D^z = - D^z_{n,c}$. 

Suppose that the system has nonzero projection onto only one eigenmode $\psi_{n,i}$ ($i = 3, \dots, 6$) at each $n$. In this case the right- and left-moving polarization vectors are related through
\begin{equation}
    \bm{P}_{R,n} = \zeta_{n,i} \bm{P}_{L,n},
\end{equation}
as seen from the eigenmode structures in Eq.~\eqref{eq:collectivemodes}. Thus $\arg(\zeta_{n,i})$ measures the phase offset between the two beams. Assume further that every occupied eigenmode is stable: $\textrm{Im}(\Omega_{n,i}) = 0$. Then $\mathrm{Im}(\zeta_{n,i}) = 0$ and consequently $\bm{P}_{R,n}$ and $\bm{P}_{L,n}$ are parallel in the sense that $\bm{P}_{L,n}^* \times \bm{P}_{R,n} = 0$. Since we assume this holds for all $n$, we conclude that $D^z$ is constant. We also see from Eq.~\eqref{eq:collectivemodes} that $\bm{P}_{R,n}$ and $\bm{P}_{L,n}$ have the same wave polarization, being either RH ($i = 3, 4$) or LH ($i = 5, 6$). Lastly, using
\begin{equation}
    \Psi_n(t) = c_{n,i}(0) \psi_{n,i} e^{-i \Omega_{n,i} t}
\end{equation}
with coefficient $c_{n,i}(0)$, we observe that the flavor waves co-precess in the sense that the phases $\varphi^a_{I,n}$ all develop at the same frequency at each order $n$.

In general, the evolution of $D^z$ is given in terms of $\Psi_n$ by
\begin{equation}
    \dot{D}^z = 8\sum_{n>0}\mathrm{Re}\!\left(\Psi_n^\dagger Z \Psi_n\right),
\end{equation}
where $Z$ is a fixed $6\times6$ matrix with all entries vanishing except for $Z_{42}=1$ and $Z_{51}=-1$. At every $n$, $\Psi_n$ is a mixture of modes $3, \dots, 6$ by our assumption of transverse ($P_{I,n}^z = 0$) flavor waves:
\begin{equation}
  \Psi_n=c_{n,3}\psi_{n,3}+c_{n,4}\psi_{n,4}+c_{n,5}\psi_{n,5}+c_{n,6}\psi_{n,6}.
\end{equation}
It follows that
\begin{equation}
\begin{split}
    \dot{D}^z=&8\mathrm{Re}\sum_{n>0}([c_{n,3}^*\psi_{n,3}^\dagger+c_{n,4}^*\psi_{n,4}^\dagger+c_{n,5}^*\psi_{n,5}^\dagger+c_{n,6}^*\psi_{n,6}^\dagger]Z[c_{n,3}\psi_{n,3}+c_{n,4}\psi_{n,4}+c_{n,5}\psi_{n,5}+c_{n,6}\psi_{n,6}])\\
    =&8\mathrm{Re}\sum_{n>0}([c_{n,3}^*\psi_{n,3}^\dagger+c_{n,4}^*\psi_{n,4}^\dagger]Z[c_{n,3}\psi_{n,3}+c_{n,4}\psi_{n,4}]+[c_{n,5}^*\psi_{n,5}^\dagger+c_{n,6}^*\psi_{n,6}^\dagger]Z[c_{n,5}\psi_{n,5}+c_{n,6}\psi_{n,6}])\\
    =&8\mathrm{Re}\sum_{n>0}([c_{n,3}+c_{n,4}]^*[c_{n,3}2i\zeta_{n,3}+c_{n,4}2i\zeta_{n,4}]+[c_{n,5}+c_{n,6}]^*[-c_{n,5}2i\zeta_{n,5}-c_{n,6}2i\zeta_{n,6}]).
\end{split}
\end{equation}
Letting
\begin{equation}
    \chi_{n,ij}\equiv\mathrm{Re}([c_{n,i}+c_{n,j}]^*[c_{n,i}2i\zeta_{n,i}+c_{n,j}2i\zeta_{n,j}])
\end{equation}
with $ij$ being either $34$ or $56$, we note that there are two possible scenarios. If $\mathrm{Im}(\Omega_{n,i})$ and $\mathrm{Im}(\Omega_{n,j})$ are both real, then $\zeta_{n,i}$ and $\zeta_{n,j}$ are also both real and
\begin{equation}
    \chi_{n,ij}=2(\zeta_{n,i}-\zeta_{n,j})[\mathrm{Re}(c_{n,i})\mathrm{Im}(c_{n,j})-\mathrm{Im}(c_{n,i})\mathrm{Re}(c_{n,j})].
    \label{eq:cross_term}
\end{equation}
On the other hand, if $\mathrm{Im}(\Omega_{n,i})$ and $\mathrm{Im}(\Omega_{n,j})$ are complex conjugates, then $\zeta_{n,i}$ and $\zeta_{n,j}$ also form a complex conjugate pair and
\begin{equation}
    \chi_{n,ij}=-2|c_{n,i}|^2\mathrm{Im}(\zeta_{n,i})-2|c_{n,j}|^2\mathrm
    {Im}(\zeta_{n,j}).
    \label{eq:direct_term}
\end{equation}
With this notation, we find that
\begin{equation}
    \dot{D}^z=8\sum_{n>0}(\chi_{n,34}-\chi_{n,56}).
\end{equation}

Consider a pair of stable eigenmodes $\psi_{n,i}$, $\psi_{n,j}$. In the stable case we find a posteriori that $D^z$ is constant apart from small fluctuations. Then the eigenmodes and $\zeta$ factors are approximately time-independent, and we have
\begin{equation}
    \chi_{n,ij} (t) = 2(\zeta_{n,i}-\zeta_{n,j}) \left[ \mathrm{Re} \left( c_{n,i}(0) e^{-i \Omega_{n,i} t} \right) \mathrm{Im} \left( c_{n,j}(0) e^{-i \Omega_{n,j} t} \right) - \mathrm{Im} \left( c_{n,i}(0) e^{-i \Omega_{n,i} t} \right) \mathrm{Re} \left( c_{n,j}(0) e^{-i \Omega_{n,j} t} \right) \right]. \label{eq:stable_chi_static}
\end{equation}
Define initial phase $\alpha_{n,i}$ via
\begin{equation}
    c_{n,i}(0) = | c_{n,i}(0)| e^{i \alpha_{n,i}}.
\end{equation}
Then 
\begin{align}
    \chi_{n,ij} (t) &= 2(\zeta_{n,i}-\zeta_{n,j}) |c_{n,i}(0)| |c_{n,j}(0)| \left[ \cos (\alpha_{n,i} - \Omega_{n,i}t) \sin (\alpha_{n,j} - \Omega_{n,j} t) - \sin (\alpha_{n,i} - \Omega_{n,i}t) \cos (\alpha_{n,j} - \Omega_{n,j} t) \right] \notag\\
    &= 2 (\zeta_{n,i}-\zeta_{n,j}) |c_{n,i}(0)| |c_{n,j}(0)| \sin \left( \alpha_{n,i} - \alpha_{n,j} - (\Omega_{n,i} - \Omega_{n,j})t \right).
\end{align}
Under time averaging (see Eqs.~\ref{eq:Omega34} and \ref{eq:Omega56}), $\chi_{n,ij}$ approximately averages to 0 unless the resonance condition $\Omega_{n,i} \approx \Omega_{n,j}$ is met. Neglecting such nonresonant terms is known as the secular approximation \cite{johns2025local}. If there are only stable nonresonant eigenmodes, then
\begin{equation}
    \overline{D^z} = \textrm{constant}
\end{equation}
under the secular approximation, reproducing Eq.~\eqref{eq:Dzoverline}. The calculation above shows that a fluctuating dynamical equilibrium can be accounted for by the averaging out of rapid phases. This is a realization of the idea that dynamical equilibrium can be maintained by effectively random subgrid fluctuations \cite{johns2023thermodynamics}.

Now consider an unstable eigenmode $\psi_{n,i}$. From Eq.~\eqref{eq:direct_term} we see that $\psi_{n,i}$ induces a change in $D^z$. If it remains unstable as $D^z$ changes, then $D^z$ changes secularly as opposed to merely fluctuating. (Fluctuations without secular change may occur if $D^z$ is marginally stable in a mixing equilibrium.) We see from Eq.~\eqref{eq:collectivemodes} that $\bm{P}_{R,n}$ and $\bm{P}_{L,n}$ have uniform RH or LH polarization. They have phase difference $\delta_n^x \neq 0, \pm\pi$ because $\zeta_{n,i}$ is complex.

\SMsection[sm:equil]{Varieties of collisionless equilibria}

In Sec.~\ref{sm:synch} we drew a distinction between dynamical equilibria and mixing equilibria. These are both notions of collisionless equilibrium, but they differ in whether or not the influence of subgrid fluctuations on coarse-grained averages is taken into account. As we define it, a dynamical equilibrium is static even under this influence. Dynamical equilibria therefore comprise a subset of mixing equilibria.

According to the analysis of Sec.~\ref{sm:eigen}, an unstable mixing equilibrium (which has at least one unstable flavor-wave eigenmode $\psi_{n,i}$) is generally not in dynamical equilibrium because the instability of $\psi_{n,i}$ drives a secular change in $D^z$. The possible exception is if the system is marginally stable, in which case the system may temporarily fluctuate into the unstable regime. In such a case, a system in mixing equilibrium may at times be unstable in the sense that some $\psi_{n,i}$ is unstable. But if the system is persistently unstable, then it is in a secularly evolving state of mixing equilibrium and not in a state of dynamical equilibrium.

The analysis in Sec.~\ref{sm:eigen} also suggests that a stable mixing equilibrium is in dynamical equilibrium provided that (1) the secular approximation holds and (2) there are no $\Omega_{n,i} \approx \Omega_{n,j}$ resonances. If the secular approximation fails, there is no apparent reason why $D^z$ could not secularly change in the absence of instability. In particular, if the system is being externally driven (\textit{e.g.}, by neutrino emission and absorption), then the secular approximation may fail because the eigenmodes are changing and thus the assumption leading to Eq.~\eqref{eq:stable_chi_static} breaks down. 

In the main text we distinguish between two types of dynamical equilibria. This distinction pertains to the dynamics of the flavor waves themselves, not to their influence on the coarse-grained averages. A \textit{quasilinear equilibrium} is a dynamical equilibrium in which the staticity of all $|c_{n,i}|$ is maintained according to the quasilinear equations of motion \cite{fiorillo2024fast}. A \textit{nonlinear equilibrium}, or a \textit{BGK mode}, is a dynamical equilibrium in which staticity of all $|c_{n,i}|$ is maintained according to the fully nonlinear equations of motion. (Staticity of all $|c_{n,i}|$ under nonlinear wave--wave interactions was termed \textit{flavor-wave equilibrium} in Ref.~\cite{johns2025local}.) We have shown that if the system has nonvanishing overlap with only a single stable flavor-wave eigenmode at each populated $n$, then the system is in quasilinear equilibrium. We have not succeeded in identifying exact BGK modes except in the single-wave case, where the stable pendulum fixed points are nonlinear equilibria. The flavor-wave pendulum of Ref.~\cite{duan2021flavor} prefigures the analysis we have given here of nonlinear Landau damping and collisionless equilibrium in single-wave FFC.


\end{document}